\documentclass{iauc}
\usepackage{graphicx}

\title[Kinematics and stellar populations of dEs] 
{Internal kinematics \& stellar populations of dE galaxies: 
Clues to their formation/evolution}

\author[Prugniel et al.]   
{Philippe Prugniel$^{2,3}$, Igor Chilingarian$^{1,2}$, Olga Sil'chenko$^3$
 \break \and Victor Afanasiev$^4$}

\affiliation{$^1$Sternberg Astronomical Institute of MSU, Russia \break 
$^2$CRAL Observatoire de Lyon, France \break 
$^3$GEPI Observatoire de Paris-Meudon, France \break 
$^4$Special Astrophysical Observatory of RAS, Russia}

\pubyear{2005}
\volume{198} 
\pagerange{1 -- 4}
\date{?? and in revised form ??}
\jname{Proceedings Near-field cosmology with dwarf elliptical galaxies }
\editors{Helmut Jerjen \& Bruno Binggelli}
\begin{document}

\maketitle

\begin{abstract}
What is the origin of the numerous population of diffuse elliptical
galaxies (dE) in clusters? These galaxies formed their stars several
billion years ago and lost their gas. Though the stellar winds resulting
from star formation and the interactions with the environment undoubedtly
play a role, their respective role and details of the mechanism of this
evolution is still debated.

In this presentation we will review the first 3D spectroscopic
observations of a handful of dE galaxies. These data reveal
complex kinematical structures, with embedded discs and counter
rotating cores, and they open extremely promising perspectives
for studying the history of the stellar population throughout
these various features. 

The presence of disks, which was already known from detailed
image analysis, and of complex kinematics and the new constraints
on the stellar population enforce the hypothesis of the evolutionary
connection between dEs and disk galaxies.


\end{abstract}

\firstsection 
\section{Introduction}
Though diffuse elliptical galaxies represent majority of the galaxy
populations in dense regions of the nearby Universe like rich clusters,
their origin and evolution remain still a matter of debate. Two main
scenarios proposed so far are (a) early hierarchical collapse with a feedback
of star formation (\cite{Dekel86}, \cite{Nagashima04}), and
(b) environmental evolution: ram pressure stripping of disc galaxies in
clusters (\cite{Mori00}) and groups (\cite{Marcolini03}), or
gravitational harassment (\cite{Moore98}). The predictions of the
popular now CDM cosmological simulations for the number and intrinsic
structures of the nearby dwarf galaxies does not agree with the real
situation, as well as the multi-burst star formation histories of
dSph of the Local Group are quite unexpected in the frame of the
hierarchical concept (\cite{Koch04}).
So observations remain the only valid way to find a clue
to their formation and evolution.

The dynamical structure of the dEs is not clear too.
From a sample of ~35 dEs, only a few are not consistent with rotational
flattening (prolate or triaxial) (\cite{Prugniel03}). However, there
are non-rotating ellipsoidal dwarfs like tidally
stressed objects (like NGC205), or some other non-rotating flat objects,
like IC~794, probably anysotropic.
Fine structures: embedded bars and discs sometimes with spiral
patterns have been found in some dEs (\cite{Jerjen00},
\cite{Barazza02}). They may argue for a hypothesis of common origin
of dEs and dIrrs.

By using the benefits of the integral-field spectroscopy
providing two-dimensional distributions of the kinematical parameters,
as well as of absorption-line indices, we aim to search for (1) kinematical
counterparts of the morphological fine structures revealed by
the surface photometry, (2) related subpopulations in the stellar
content.

\section{Observations, data reduction and analysis}
In 2004 we started to observe dE galaxies in Virgo cluster with the MPFS
integral field unit (IFU) spectrograph at the Russian 6-m telescope.
The technique of 3D spectroscopy allows to obtain spatially
resolved distribution of kinematical characteristics (radial velocity and
velocity dispersion fields) independently for gas and stars, and stellar
population parameters (for example, maps of line-strength indices)
in a single exposure.

Observations of the galaxies have been performed with the MPFS IFU
spectrograph on the 6m telescope of SAO RAS during three observing runs in
2004. The parameters of the observations are presented in Tab.~1

\begin{table}
\centering
\begin{tabular}{|l|l|r|r|c|r|}
\hline
Name & Date & seeing & $t_{exp}$ & S/N$_{cent}$ & $\sigma_{cent}$ \\
\hline
IC~3468 & 2004/Mar/20 & 1.5'' & 2.5h & 20 & 30 \\
IC~3653 & 2004/May/24 & 1.4'' & 2h & 35 & 70 \\
NGC~770 & 2004/Oct/07 & 2'' & 2h & 50 & 100 \\
\hline
\end{tabular}
\caption{Parameters of observations}
\end{table}

Data processing and analysis are discussed by
Chilingarian et al. (this volume). We refer there paper for 
explanations of adaptive binning techinque and fitting procedures.

\begin{figure}
\includegraphics[width=13cm,height=4.5cm]{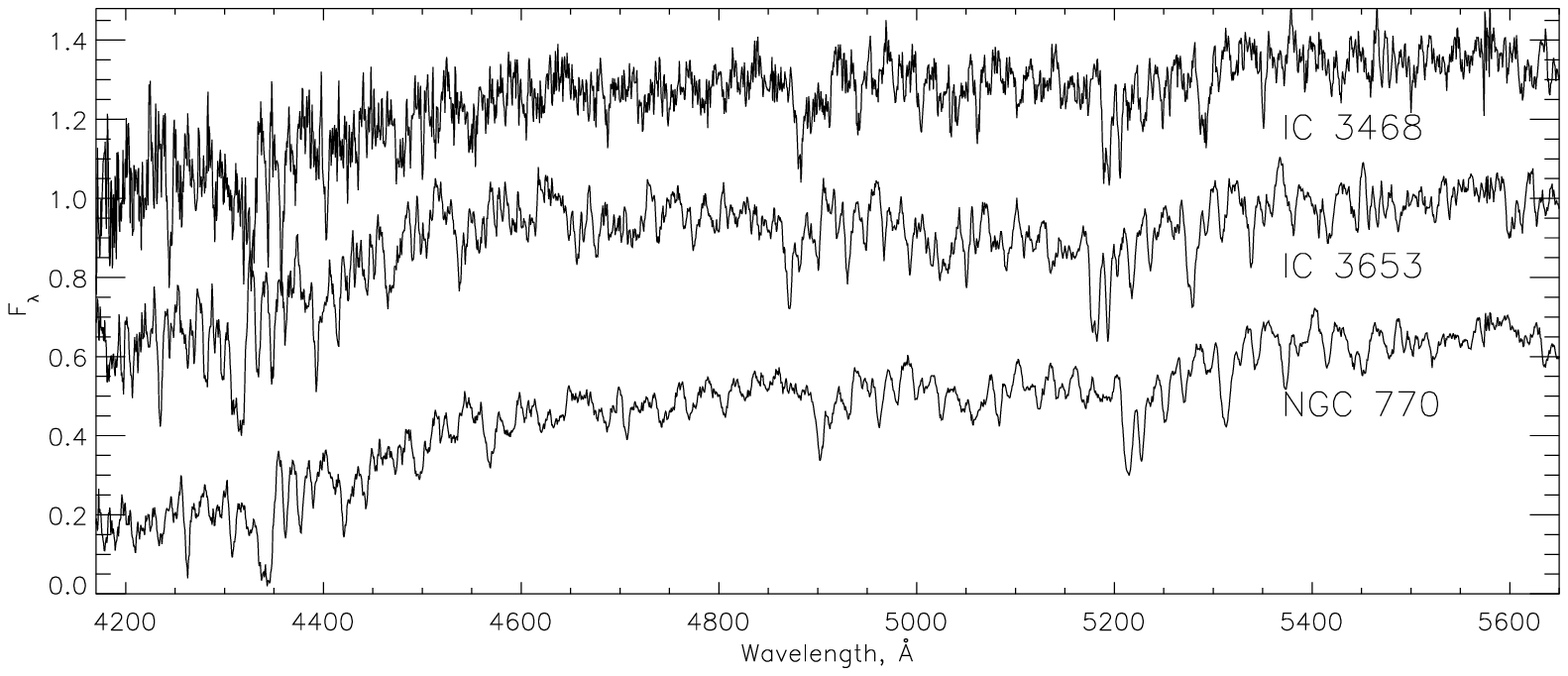}
\begin{tabular}{l l l l}
 \hline
 (a) IC~3468 \\
 \includegraphics[width=3.2cm]{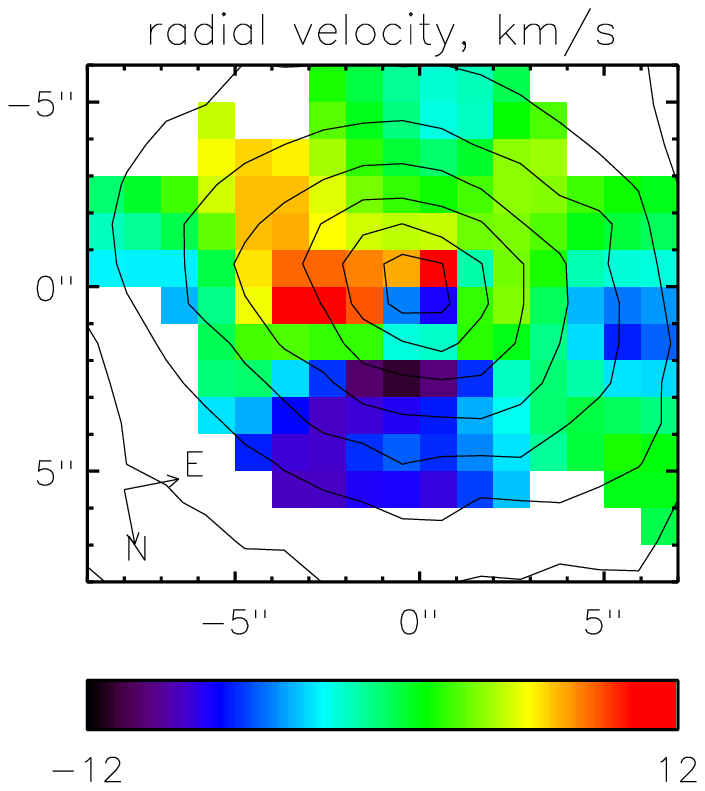} &
 \includegraphics[width=3.2cm]{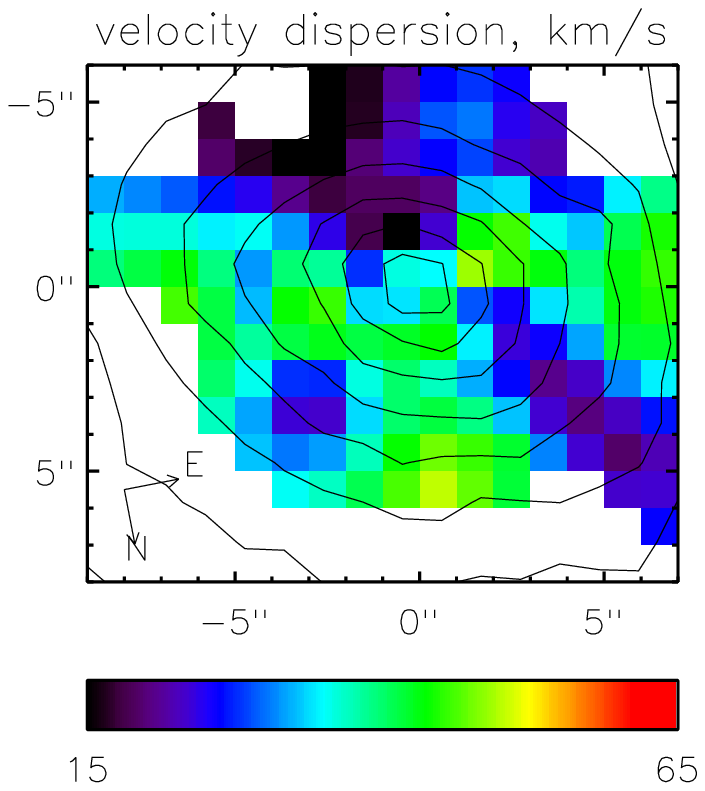} &
 \includegraphics[width=3.2cm]{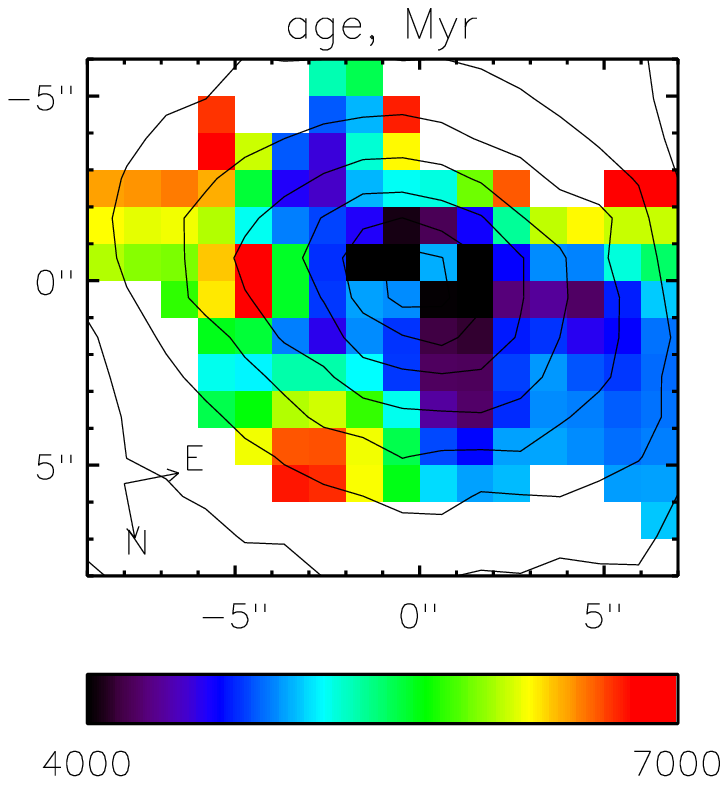} &
 \includegraphics[width=3.2cm]{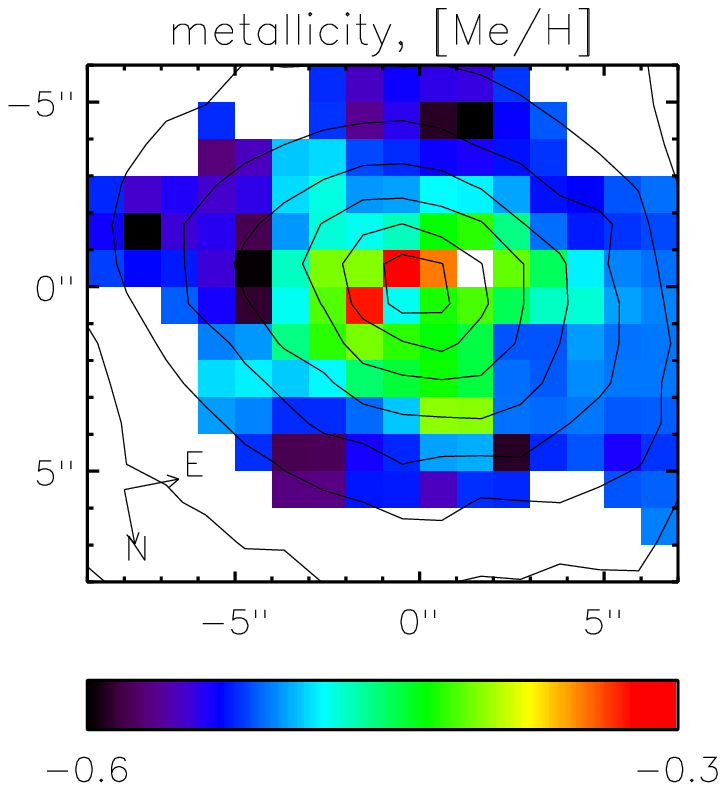} \\
 \hline
 (b) IC~3653 \\
 \includegraphics[width=3.2cm]{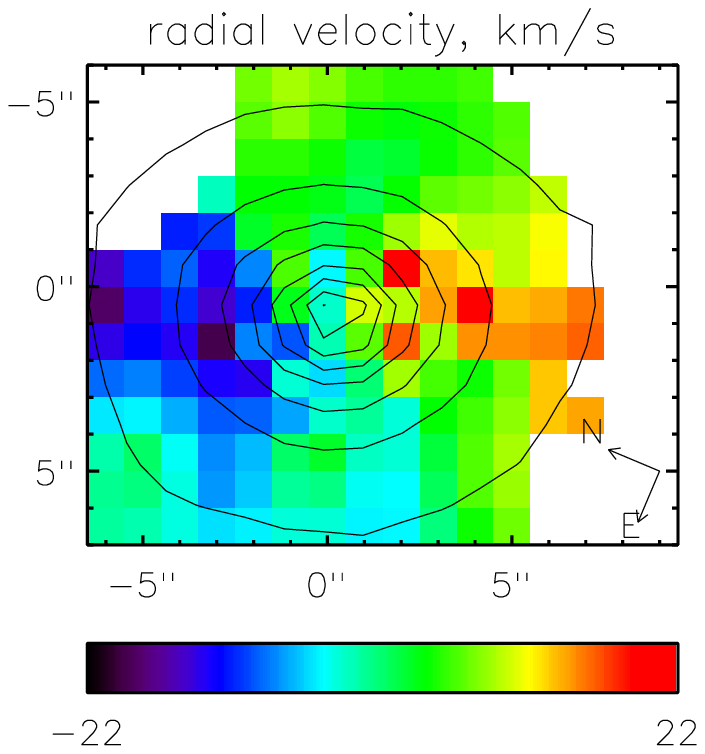} &
 \includegraphics[width=3.2cm]{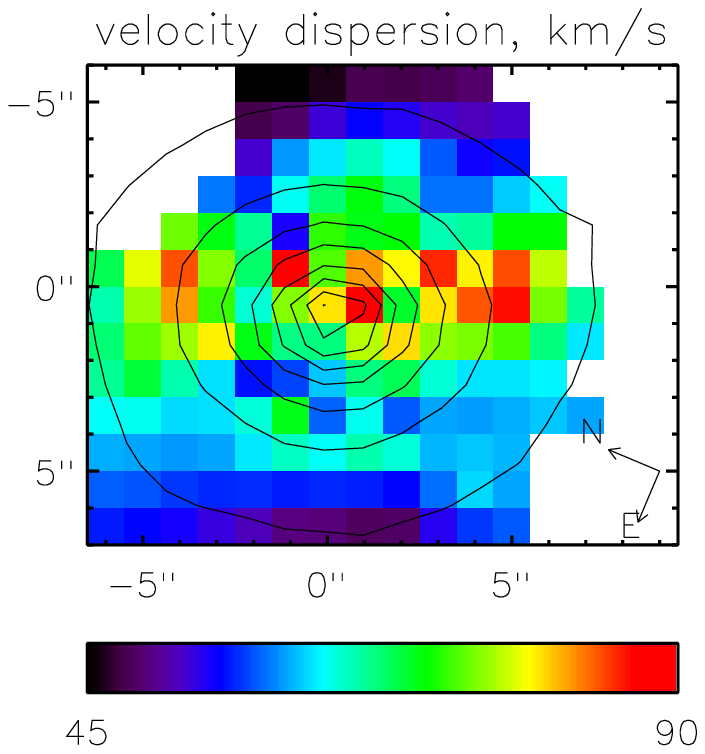} &
 \includegraphics[width=3.2cm]{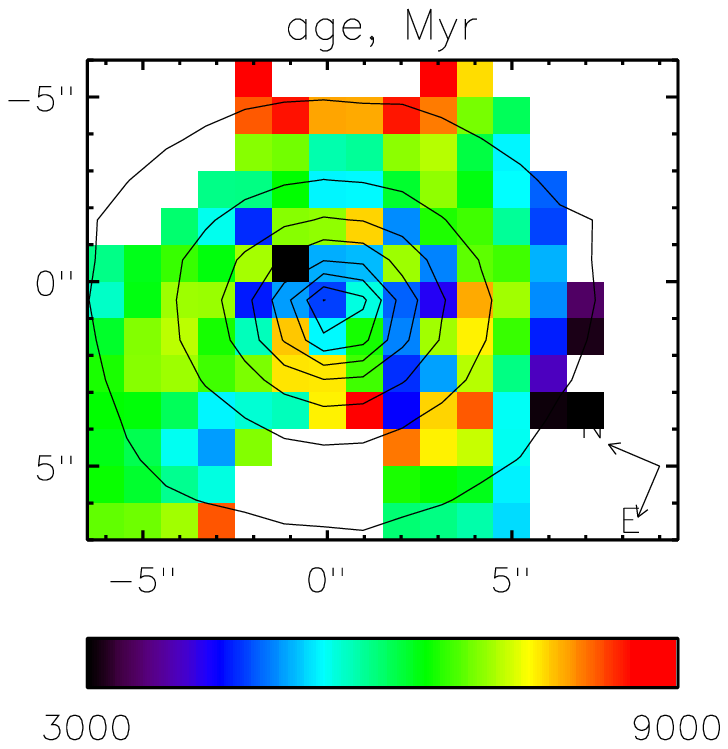} &
 \includegraphics[width=3.2cm]{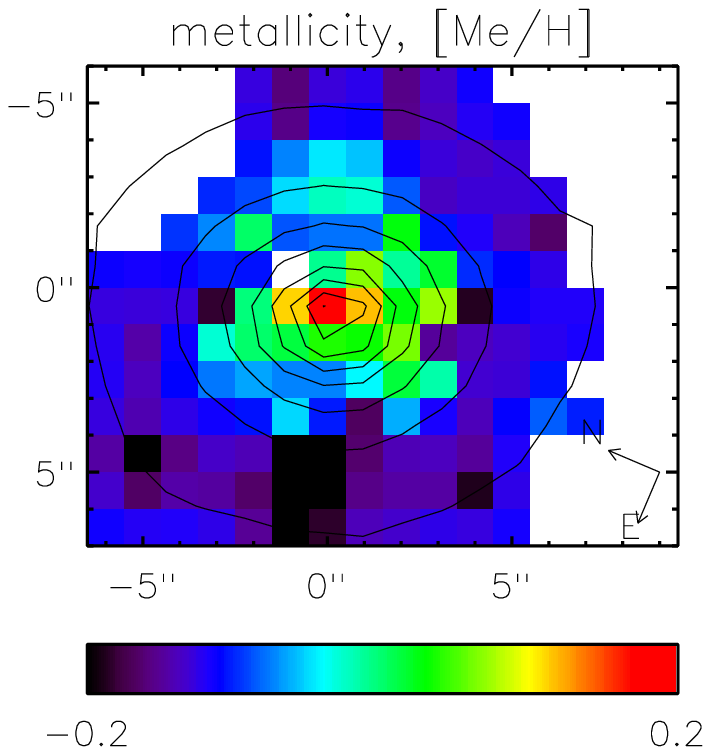} \\
 \hline
 (c) NGC~770 \\
 \includegraphics[width=3.2cm]{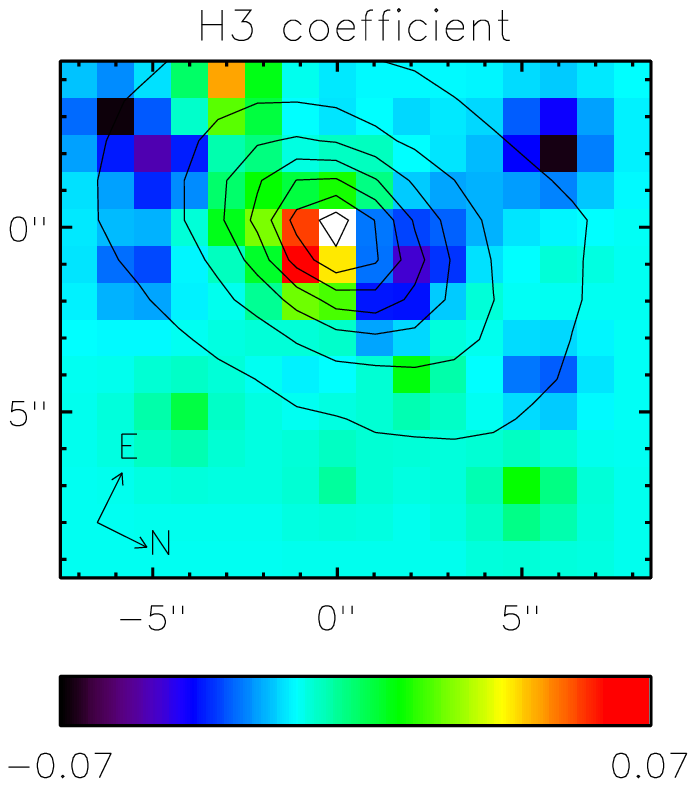} &
 \includegraphics[width=3.2cm]{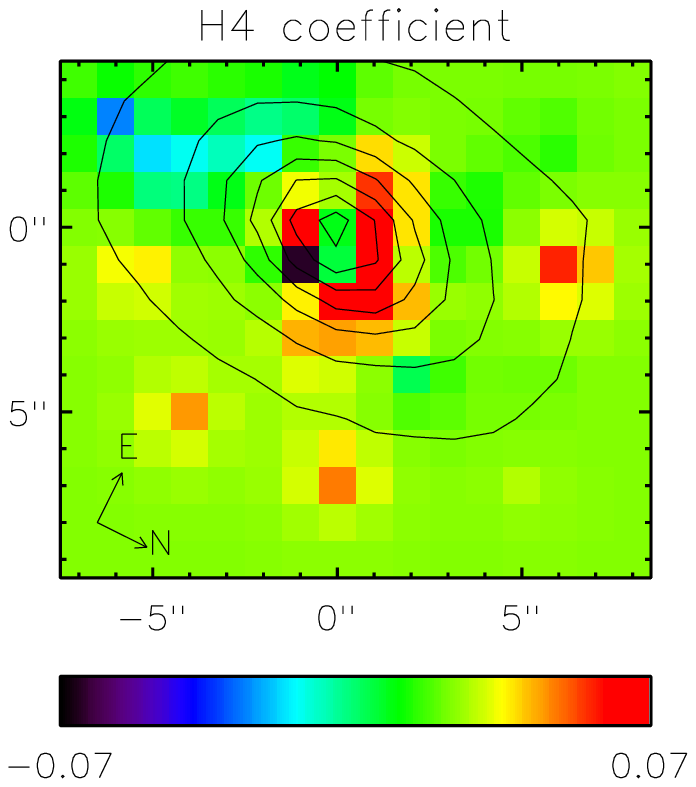} &
 \includegraphics[width=3.2cm]{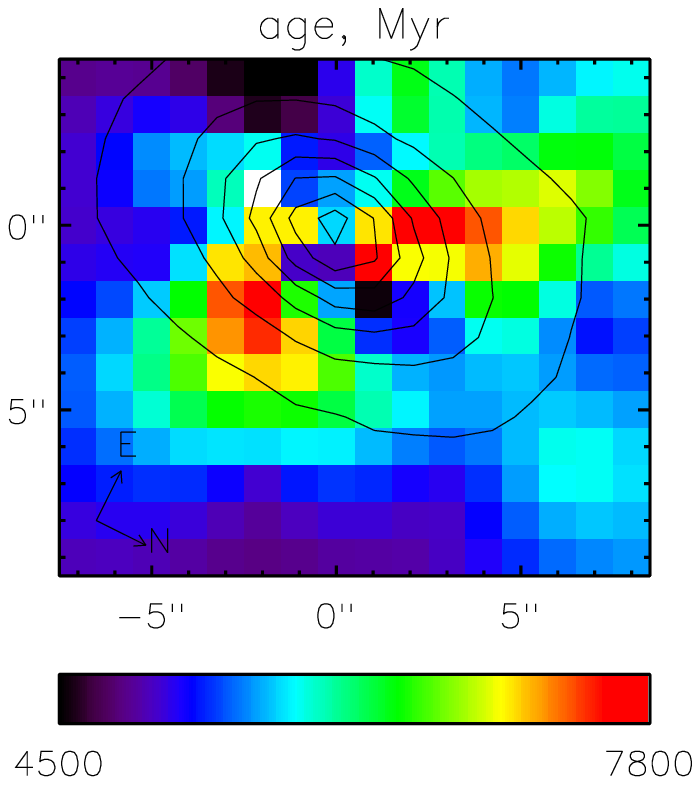} &
 \includegraphics[width=3.2cm]{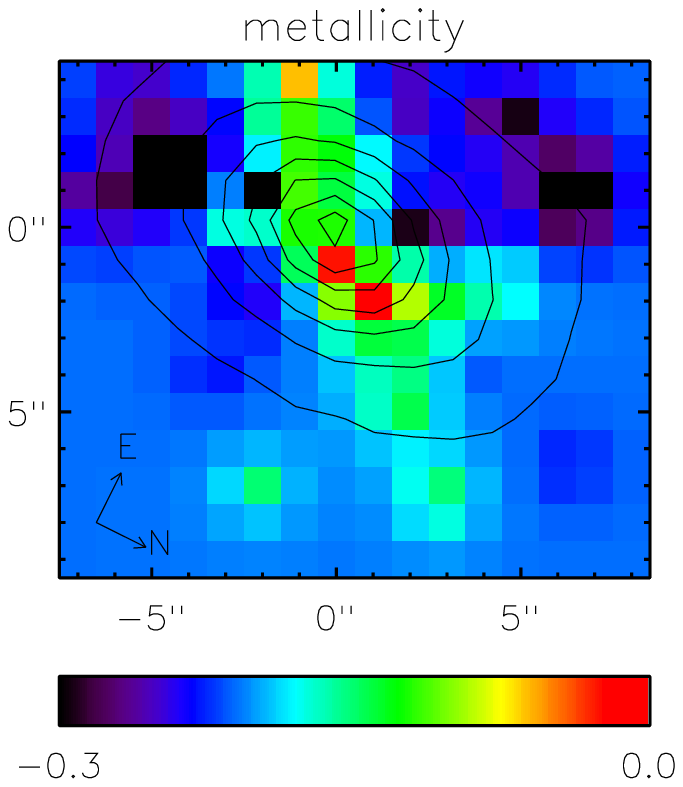} \\
\end{tabular}
\begin{tabular}{l l l}
 \includegraphics[width=3.2cm]{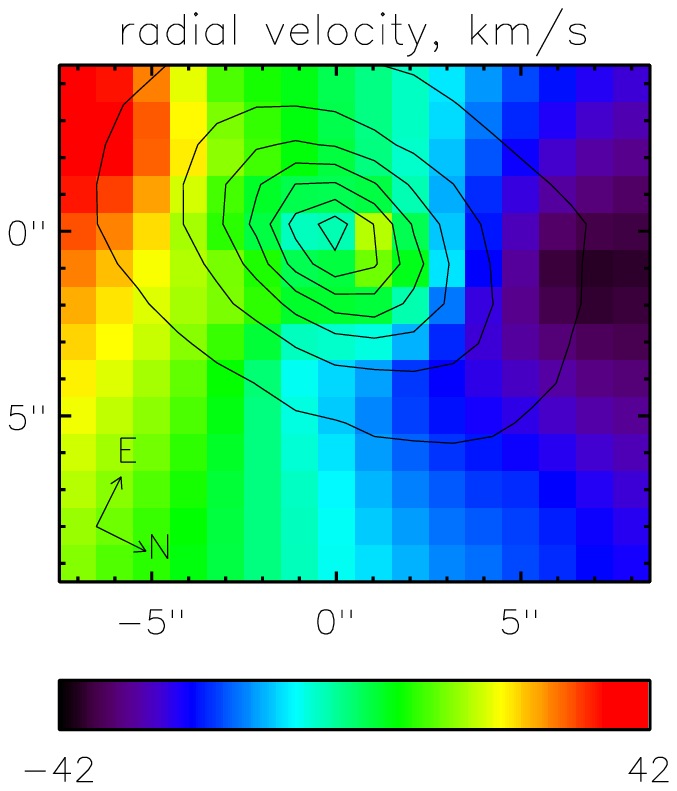} &
 \includegraphics[width=3.2cm]{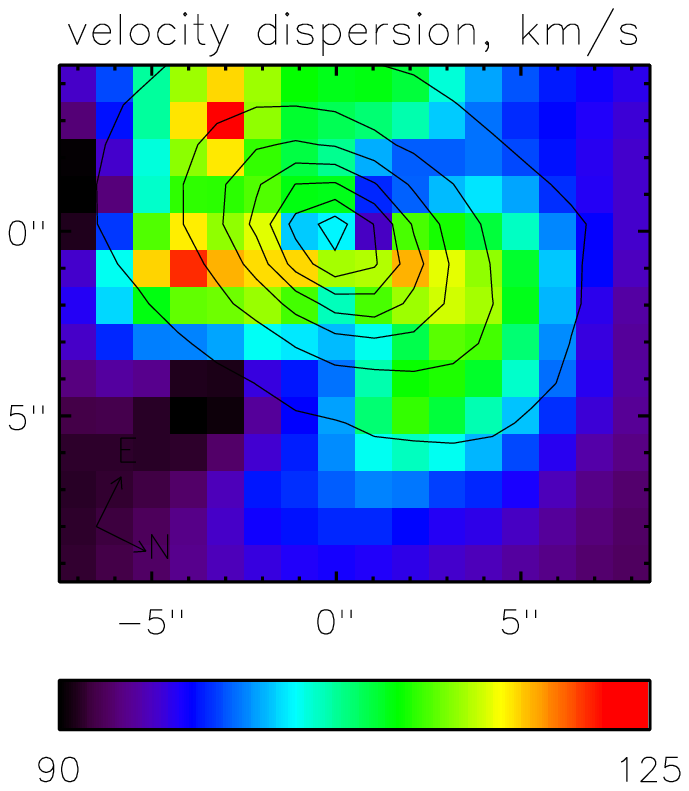} &
 \includegraphics[width=6.2cm,height=3.3cm]{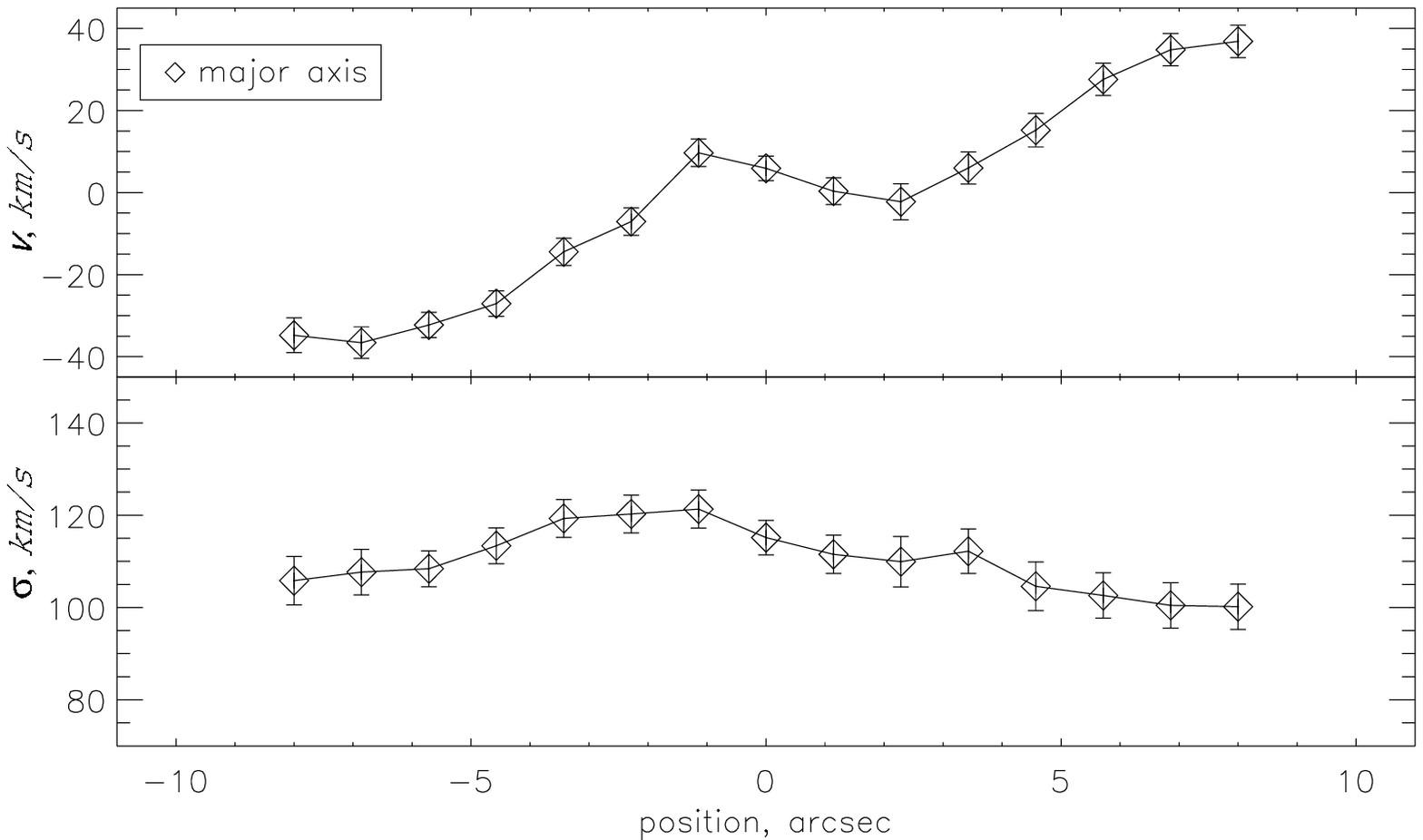} \\
\end{tabular}
\caption{On the upper figure spectra of the central regions of
 three galaxies are shown. (a) and (b) Results for IC~3468 and IC~3653:
 2D fields of radial velocity, velocity dispersion, age and metallicity
 in the single burst SFH model. (c) Results for NGC~770: the same as previous,
 besides the maps of H3 and H4 Gauss-Hermite coefficients, and kinematical
 profiles along major axis are shown.}
\end{figure}

\section{Results}
Here we present the results for three diffuse galaxies: IC3468 and IC3653
in the Virgo cluster, and NGC770 in the NGC772 group. We are modeling the
simplest star formation history containing one burst.

\subsection{IC3468}
Rather low S/N ratio doesn't allow very precise measurements. The
results shown are obtained for tessellations with target S/N=10. No young
subpopulation is detected. For a main burst with $t=5$ Gyr with low metallicity
([m/H]=-0.4) we see some metallicity gradient (from -0.3 in the centre till
-0.55 in the outskirts). Velocity dispersion map shows stripe with low
values (near the limit of the measurements, about 20 km/s) aligned along the
major axis of the isophotes. Radial velocity field is quite irregular and
asymmetric, though showing some kind of rotation with the amplitude of 10 km/s.

\subsection{IC3653}
Good S/N ratio allows to build precise maps of the parameters till 7'' from
the centre of the galaxy. Disc rotation with the amplitude of about 40 km/s
is seen in the radial velocity field with the strongest gradient in the
inner region, that is quite unusual for objects of such type. Velocity
dispersion shows gradient from 90 km/s in the centre till 45 km/s in the
outer regions. Age map is almost flat with a mean age of 7 Gyr, all details
seen are not statistically significant except young values in the very
centre. Metallicity distribution shows shallow gradient from $[m/H]=0$ in
the centre to $[m/H]=-0.2$ at the outskirts with a small metal-rich nuclear
region ([m/H]=0.2). There is a marginal detection of younger
population $t \sim 1$ Gyr in the centre which may contain a mass fraction
of 1 to 3 percent. The maps are computed for target S/N values of 15.

\subsection{NGC770}
The galaxy experiences strong tidal interaction with the giant spiral
NGC772. Its luminosity (M$_B$=-18.4) and velocity dispersion (110-120 km/s)
place it between giants and dwarfs. Very good S/N ratio allows precise
measurements of all the parameters (for example, $\Delta v \sim 1 km/s$). 
Good sampling of LOSVD due to high velocity dispersion allows to measure H3
and H4 Gauss-Hermit coefficients. We see impressive kinematically decoupled
core in this object. There are evidences for considering it to be a
counter-rotating young metal-rich highly inclined stellar disc aligned
almost with the major axis of the galaxy: (1) velocity dispersion map shows
larger values on the "switches" of the velocity, there is a stripe with
lower values orthogonal to the structures in age/metallicity distributions;
(2) map of H3 coefficient shows regions with positive and negative values
before and after "switch" of the velocity; (3) age and metallicity
distributions clearly show the decoupled structure aligned almost along the
major axis having younger age ($t \sim 4.5 Gyr$ vs $7 Gyr$) and higher
metallicity ($[Me/H] \sim -0.05$ vs $-0.2$) than the surrounding spheroid.
The galaxy probably had a partially dissipative merger event, that induced a
burst of star formation some 4.5 Gyr ago producing the central stellar
structure that we see now as a counter rotating subpopulation with the high
metallicity.

\section{Conclusions}
Kinematical features found in the dE galaxies studies strengthen the
connection between dEs and dIrrs. Inner discs are probably the remnants
of the stellar discs existing before gas removal.

Whatever the gas-removal mechanism is (winds or ram pressure stripping) the
signatures would be the same. But these systems probably did not undergo
strong relaxation which would have erased the structures and population
gradients. Most dEs should have the same origin, i.e. the same type of
progenitors (pre-dI). They had a different history, i.e. interplay between
feedback, ram pressure stripping and tidal effects. The investigation
of the properties of the kinematical and
population substructures will help to understand this history.

\begin{acknowledgments}
We would like to thank International Astronomical Union providing financial
support for attending IAU Colloquium 198, 6-m telescope time allocation
committee, INTAS foundation providing PhD fellowship for I.C.,
EGIDE and CNRS for additional support.
\end{acknowledgments}

\end{document}